\documentclass[notitlepage,nofootinbib,preprintnumbers,amssymb,superscriptaddress]{revtex4-1}

\usepackage{amsfonts,amssymb,amsmath,graphicx,color,bm}
\definecolor{ultramarine}{rgb}{0.07, 0.04, 0.56}
\definecolor{cadmiumgreen}{rgb}{0.0, 0.42, 0.24}
\definecolor{indigo(dye)}{rgb}{0.0, 0.25, 0.42}
\usepackage[linktocpage=true]{hyperref}
\hypersetup{
colorlinks=true,
citecolor=ultramarine,
linkcolor=cadmiumgreen,
urlcolor=indigo(dye),
}

\newcommand{\fr}[2]{\frac{#1}{#2}}
\newcommand{\pa}{\partial}
\newcommand{\ti}{\tilde}
\newcommand{\na}{\nabla}
\newcommand{\bra}[1]{\left( #1 \right)}  
\newcommand{\brb}[1]{\left[ #1 \right]}  
  
\newcommand{\be}{\begin{equation}}  
\newcommand{\ee}{\end{equation}}
\newcommand{\bem}{\begin{bmatrix}}
\newcommand{\eem}{\end{bmatrix}}
\newcommand{\Mpl}{M_{\rm Pl}}

\newcommand{\ga}{\gamma}

\newcommand{\la}{\lambda}
\newcommand{\si}{\sigma}

\newcommand{\mn}{\mu \nu}
\newcommand{\mE}{\mathcal{E}}
\newcommand{\mJ}{\mathcal{J}}
\newcommand{\mF}{\mathcal{F}}
\newcommand{\mG}{\mathcal{G}}
\newcommand{\mH}{\mathcal{H}}
\newcommand{\mL}{\mathcal{L}}

\begin{document}

\preprint{RESCEU-30/16}

\title{Linear perturbation analysis of hairy black holes in shift-symmetric Horndeski theories: Odd-parity perturbations}

\author{Kazufumi Takahashi}
\affiliation{Research Center for the Early Universe (RESCEU),
Graduate School of Science, The University of Tokyo, Tokyo 113-0033, Japan}
\affiliation{Department of Physics, Graduate School of Science, 
The University of Tokyo, Tokyo 113-0033, Japan}
\author{Teruaki Suyama}
\affiliation{Research Center for the Early Universe (RESCEU),
Graduate School of Science, The University of Tokyo, Tokyo 113-0033, Japan}

\begin{abstract}
We analyze the mode stability of odd-parity perturbations of black holes with linearly time-dependent scalar hair in shift-symmetric Horndeski theories.
We show that a large class of black hole solutions in these theories suffer from ghost or gradient instability, while there are some classes of solutions that are stable under linear odd-parity perturbations in the context of mode analysis.
\end{abstract}

\maketitle
\section{Introduction}

There is a growing interest in healthy scalar-tensor theories of gravity, among which the so-called Horndeski or generalized Galileon class~\cite{Horndeski:1974wa,Deffayet:2011gz,Kobayashi:2011nu} has been studied extensively.
This class has field equations which contain at most second-order derivatives for both the metric and the scalar field.
Such a nature is desirable as, in general, higher derivatives in equations of motion (EOMs) yield Ostrogradsky ghosts~\cite{Woodard:2015zca}.
The Lagrangian of the Horndeski theories consists of four parts $\mL_i~(i=2,3,4,5)$, each of which is characterized by an arbitrary function $G_i(\phi,X)$ that depends on the scalar field $\phi$ and its canonical kinetic term $X$.

Within the context of the Horndeski theories, it is important to investigate what kind of black hole (BH) solution exists.
Since the additional scalar degree of freedom (DOF) is introduced, one naturally expects that BHs could support scalar hair \cite{Sotiriou:2015pka,Babichev:2016rlq} other than mass, angular momentum, and electrical charge.
Regarding this point, an interesting subclass of the Horndeski theories is the one possessing shift symmetry of the scalar field, i.e., such that the four arbitrary functions do not depend on $\phi$ explicitly:~$G_i(X)$.
Although there is a no-hair theorem for static spherically symmetric BHs in this class~\cite{Hui:2012qt}, one can circumvent the prohibition by introducing a linearly time-dependent scalar field~\cite{Babichev:2013cya}.\footnote{For other loopholes of the no-hair theorem in \cite{Hui:2012qt}, see \cite{Sotiriou:2013qea} or reviews \cite{Herdeiro:2015waa,Silva:2016smx}.}
This linear time dependence does not contradict the static ansatz for the metric by virtue of the shift symmetry.
Such a possibility for BH solutions in the shift-symmetric Horndeski theories has been explored in recent works~\cite{Babichev:2013cya,Kobayashi:2014eva,Babichev:2016fbg}.
Among these, \cite{Babichev:2013cya,Kobayashi:2014eva} found solutions of this type under a restrictive situation where the action has reflection symmetry of the scalar field as well as the shift symmetry; i.e., they took into account only $\mL_2$ and $\mL_4$ that include even numbers of $\phi$.
These solutions include the Schwarzschild-de~Sitter metric with a nontrivial scalar profile, which is called {\it stealth BH} since such a solution cannot be distinguished from the one in general relativity (GR) at least at the background level.
On the other hand, \cite{Babichev:2016fbg} discovered BH solutions including $\mL_3$ term which was neglected in \cite{Babichev:2013cya,Kobayashi:2014eva}.
For these solutions, the behavior of the spacetime is different from that in GR.

Another important thing is to analyze stability of such BH solutions.
In the context of BH perturbation theory~\cite{Regge:1957td,Zerilli:1970se}, it was shown in \cite{Ogawa:2015pea} that odd-parity perturbations of a large class of BHs in the shift- and reflection-symmetric Horndeski theories are plagued by ghost or gradient instability.
As a consequence, it turned out that all the solutions given in \cite{Babichev:2013cya,Kobayashi:2014eva} are not allowed physically.
However, as we mentioned, the solutions found in \cite{Babichev:2016fbg} do not belong to the reflection-symmetric subclass, and thus, the stability analysis remains to be done.
In light of this situation, we perform the similar analysis as in \cite{Ogawa:2015pea} by including all the terms $\mL_i$.
Although there are only few known BH solutions, such a general stability analysis is meaningful:~If one could show that a certain class of solutions always suffers from some instability, then it can reduce the space of physically possible BHs that we have to survey.
This is indeed the case as is shown later, while there is some loophole.

This paper is organized as follows.
In \S \ref{hairyBH}, we introduce the shift-symmetric Horndeski theories and briefly review the hairy BH solutions discovered so far.
Then in \S \ref{opert}, the mode stability of odd-parity perturbations of the BH solutions is analyzed.
We show that a wide class of the BHs is always plagued by ghost or gradient instability as in \cite{Ogawa:2015pea}, while some narrower classes are fully stable under odd-parity perturbations for any fixed mode.
Finally, we draw our conclusions in \S \ref{conclusion}.

\section{Hairy black holes in shift-symmetric Horndeski theories}\label{hairyBH}
Throughout the present paper, we work in the following subclass of the Horndeski theory:
	\be
	S=\int d^4x\sqrt{-g}\sum_{i=2}^5\mL_i, \label{shiftsymact}
	\ee
where
	\be
	\begin{split}
	\mL_2&=G_2(X),\\
	\mL_3&=-G_3(X)\Box\phi,\\
	\mL_4&=G_4(X)R+G_{4X}\brb{(\Box\phi)^2-(\na_\mu\na_\nu\phi)^2},\\
	\mL_5&=G_5(X)G^{\mn}\na_\mu\na_\nu\phi-\fr{1}{6}G_{5X}\brb{(\Box\phi)^3-3\Box\phi(\na_\mu\na_\nu\phi)^2+2(\na_\mu\na_\nu\phi)^3}.
	\end{split}
	\ee
Here $G_i$ are arbitrary functions of the canonical kinetic term of the scalar field~$X\equiv -(\pa_\mu\phi)^2/2$, and
	\be
	(\na_\mu\na_\nu\phi)^n\equiv \na^{\mu_1}\na_{\mu_2}\phi\na^{\mu_2}\na_{\mu_3}\phi\cdots\na^{\mu_n}\na_{\mu_1}\phi,~~~(n\ge 2).
	\ee
Note that the scalar field appears only with derivatives, so the action is invariant under the shift~$\phi\to\phi+c$, where $c$ is a constant.
Since we are interested in static spherically symmetric BH solutions, the background metric is assumed to be of the form
	\be
	\bar{g}_{\mn}dx^\mu dx^\nu=-A(r)dt^2+\fr{dr^2}{B(r)}+2C(r)dtdr+D(r)r^2\ga_{ab}dx^adx^b, \label{sphsymmetric}
	\ee
where $a,b$ denote angular variables and $\ga_{ab}$ represents the metric on a two-dimensional sphere.
As for the scalar field, we assume
	\be
	\phi(t,r)=qt+\psi(r), \label{lintimedep}
	\ee
where the scalar velocity $q$ is constant.
Such a linearly time-dependent scalar configuration is compatible with the static ansatz of the metric due to the shift symmetry of the action~\eqref{shiftsymact}.
For this configuration of the scalar field, the kinetic term becomes
	\be
	X=\fr{q^2}{2A}-\fr{1}{2}B\psi'{}^2. \label{ckt}
	\ee

Now, we have five unknown variables~$A,B,C,D$, and $\psi$.
Substituting the ansatzes~\eqref{sphsymmetric} and \eqref{lintimedep} into the action~\eqref{shiftsymact} and variation with respect to the five variables yield the EOMs~$\mE_\Phi=0\,(\Phi=A,B,C,D,\psi)$ (for the detailed expressions of $\mE_\Phi$, see Appendix~\ref{appA}).
However, two of the five variables are actually unphysical DOFs which can be eliminated by using the gauge DOFs.
As usual, we set $C=0$ and $D=1$ after deriving EOMs, so the dynamical variables are only $A,B,\psi$.\footnote{While $D=1$ alone is a complete gauge fixing and thus can be imposed at the level of action, $C=0$ should be substituted only after deriving the background EOMs~\cite{Motohashi:2016prk}.}
Correspondingly, only three out of the five EOMs are independent:~$\mE_D$ and $\mE_\psi$ can be written in terms of the other components of EOMs (see Appendix~\ref{appA}).
Thus, we only need the remaining EOMs~$\mE_A=\mE_B=\mE_C=0$.

In the class of shift-symmetric Horndeski theories, some BH solutions with linearly time-dependent scalar hair have been found recently.
The solutions in \cite{Babichev:2013cya,Kobayashi:2014eva} were obtained under a restrictive situation where the action is invariant under the reflection $\phi\to-\phi$; i.e., $\mL_3$ and $\mL_5$ were neglected.
They include a solution which exactly coincides with the GR solution having a nontrivial scalar configuration, dubbed {\it stealth BH}.
On the other hand, \cite{Babichev:2016fbg} took into account $\mL_3$ and found new solutions numerically.
These solutions behave in a different way from GR, while they asymptote to de~Sitter spacetime.

Before closing this section, let us remark on the behavior of the scalar field and its kinetic term~\eqref{ckt} in the vicinity of the horizon $r=r_h$.
Although the expression~\eqref{ckt} seems to be divergent for $A\simeq0$, it is known that $X$ takes a finite value at the horizon for the solutions found in \cite{Babichev:2013cya,Kobayashi:2014eva}.\footnote{For the numerically obtained solutions in \cite{Babichev:2016fbg}, the regularity of $X$ at the horizon can be shown by use of the background EOMs.}
To cancel out the divergence of the first term in \eqref{ckt}, the radial part of the scalar field must take the form of
	\be
	\psi(r)\approx\pm q\int^r \fr{dr}{\sqrt{AB}}, \label{psihor}
	\ee
near $r=r_h$.
Let us focus on the plus branch of \eqref{psihor} and discuss the near-horizon behavior of the scalar field~\cite{Babichev:2013cya,Kobayashi:2014eva}.
If one writes $\phi(t,r)$ in terms of the ingoing Eddington-Finkelstein time coordinate
	\be
	u\equiv t+\int^r \fr{dr}{\sqrt{AB}}
	\ee
instead of $t$, the divergent part in $\psi(r)$ can be absorbed into $u$.
This means that any probe infalling to the BH observes only a finite value of the scalar field.
Therefore, in what follows, we only deal with the plus branch of \eqref{psihor}.

\section{Odd-parity perturbations}\label{opert}
Any metric perturbation $h_{\mn}\equiv g_{\mn}-\bar{g}_{\mn}$ can be separated into two parts:~the odd- and the even-parity perturbations.
We focus on the former in the present paper, while the latter will be examined in a subsequent paper since the analysis of even modes is technically involved.
The odd-parity perturbations are also known as vector-type perturbations, which can be decomposed as follows~\cite{Regge:1957td}:
	\be
	\begin{split}
	h_{tt}&=h_{tr}=h_{rr}=0,\\
	h_{ta}&=\sum_{\ell,m}h_{0,\ell m}(t,r)E_a{}^b\bar{\na}_bY_{\ell m}(\theta,\varphi),\\
	h_{ra}&=\sum_{\ell,m}h_{1,\ell m}(t,r)E_a{}^b\bar{\na}_bY_{\ell m}(\theta,\varphi),\\
	h_{ab}&=\sum_{\ell,m}h_{2,\ell m}(t,r)E_{(a}{}^c\bar{\na}_{|c|}\bar{\na}_{b)}Y_{\ell m}(\theta,\varphi),
	\end{split} \label{harmexp}
	\ee
where $E_{ab}$ is the completely antisymmetric tensor defined on a two-dimensional sphere, and $\bar{\na}_a$ denotes a covariant derivative with respect to $\gamma_{ab}$.
Since modes with different $(\ell,m)$ evolve independently, in the following we focus on a specific mode and omit the indices $\ell,m$ unless necessary.
Note that the odd-parity perturbations do not have the monopole~($\ell=0$) term, and $h_2$ is vanishing for the dipole~($\ell=1$) terms.
The perturbation of the scalar field is not taken into account as it belongs to the even-parity perturbations.

The expansion coefficients $h_0,h_1,h_2$ are not all physical DOFs, as there exist gauge DOFs corresponding to the general covariance.
The general infinitesimal transformation of coordinates for the odd modes can be written as
	\be
	x^a\rightarrow x^a+\xi^a,~~~\xi_a=\sum_{\ell,m}\Lambda_{\ell m}(t,r)E_a{}^b\bar{\na}_bY_{\ell m}(\theta,\varphi).
	\ee
Correspondingly, the coefficients $h_0,h_1,h_2$ transform as
	\be
	\begin{split}
	h_0&\rightarrow h_0-\dot{\Lambda}, \\
	h_1&\rightarrow h_1-\Lambda'+\fr{2}{r}\Lambda, \\
	h_2&\rightarrow h_2-2\Lambda.
	\end{split} \label{gaugetrnsf}
	\ee
Therefore, in the case of $\ell\ge 2$ where $h_2$ is nontrivial, one can choose $\Lambda=h_2/2$ to redefine $h_2=0$, which is a complete gauge fixing.
For the dipole modes where $h_2$ is absent, this gauge function $\Lambda$ is used to cancel out another unphysical DOF.

In what follows, we investigate $\ell\ge 2$ and $\ell=1$ modes separately and discuss the stability of BH solutions with linearly time-dependent scalar hair.

\subsection{The $\ell\ge 2$ modes}\label{opertmulti}
First, we consider modes with $\ell\ge2$.
As we mentioned, we focus on a specific mode $(\ell,m)$.
Moreover, one is allowed to set $m=0$ from the beginning, since all the terms with the same multipole~$\ell$ contributes equally by virtue of the spherical symmetry of the background.\footnote{Hence, it is more useful to expand metric perturbations in terms of the Legendre polynomials instead of the spherical harmonics. In the subsequent analysis, $h_0,h_1,h_2$ denote the coefficients of $P_\ell(\cos \theta)$.}
After performing the integration over angular variables, the second-order action
	\be
	S^{(2)}=\int dtdr\mL^{(2)}
	\ee
takes the form of
	\be
	\fr{2\ell+1}{2\pi}\mL^{(2)}=a_1h_0^2+a_2h_1^2+a_3\bra{\dot{h}_1^2-2h'_0\dot{h}_1+h'_0{}^2+\fr{4h_0\dot{h}_1}{r}}+a_4h_0h_1. \label{Lag1}
	\ee
Here dots and primes denote derivatives with respect to $t$ and $r$, respectively.
The coefficients can be written as
	\be
	\begin{split}
	a_1&=\fr{\ell(\ell+1)}{r^2}\brb{\fr{d}{dr}\bra{r\sqrt{\fr{B}{A}}\mH}+\fr{(\ell-1)(\ell+2)}{2\sqrt{AB}}\mF},\\
	a_2&=-\fr{(\ell-1)\ell(\ell+1)(\ell+2)}{2}\fr{\sqrt{AB}}{r^2}\mG,\\
	a_3&=\fr{\ell(\ell+1)}{2}\sqrt{\fr{B}{A}}\mH,\\
	a_4&=\fr{(\ell-1)\ell(\ell+1)(\ell+2)}{r^2}\sqrt{\fr{B}{A}}\mJ,
	\end{split} \label{acoeff}
	\ee
where we have used the background EOMs for the simplification.
The functions $\mF,\mG,\mH,\mJ$ are defined by\footnote{Here, $\psi'$ in the denominator of $\mF$ or $\mG$ cancels out when \eqref{ckt} is used. Such an expression was chosen to clarify terms that are divergent near the BH horizon where $A\simeq0$ [see \eqref{fghor}].}
	\be
	\begin{split}
	\mF&\equiv 2\bra{G_4-\fr{q^2}{A}G_{4X}}-\bra{\fr{q^2}{A}\fr{A'}{A}+2X'}\fr{X}{\psi'}G_{5X},\\
	\mG&\equiv 2\brb{G_4+\bra{\fr{q^2}{A}-2X}G_{4X}}+\fr{A'}{A}\bra{\fr{q^2}{A}-2X}\fr{X}{\psi'}G_{5X},\\
	\mH&\equiv 2(G_4-2XG_{4X})+\fr{2}{r}B\psi'XG_{5X},\\
	\mJ&\equiv q\bra{2\psi'G_{4X}+\fr{A'}{A}XG_{5X}}.
	\end{split} \label{fghj}
	\ee

Since the structure of the Lagrangian completely coincides with the one in \cite{Ogawa:2015pea}, the subsequent analysis proceeds in a parallel manner.
Integrating by parts, one can rewrite \eqref{Lag1} as
	\be
	\fr{2\ell+1}{2\pi}\mL^{(2)}=\bra{a_1-\fr{2(ra_3)'}{r^2}}h_0^2+a_2h_1^2+a_3\bra{\dot{h}_1-h_0'+\fr{2}{r}h_0}^2+a_4h_0h_1. \label{Lag2}
	\ee
Then, we introduce an auxiliary variable $\chi$ to write \cite{DeFelice:2011ka}
	\be
	\fr{2\ell+1}{2\pi}\mL^{(2)}=\bra{a_1-\fr{2(ra_3)'}{r^2}}h_0^2+a_2h_1^2+a_3\brb{-\chi^2+2\chi\bra{\dot{h}_1-h_0'+\fr{2}{r}h_0}}+a_4h_0h_1. \label{Lag3}
	\ee
Note that the EOM for $\chi$ yields $\chi=\dot{h}_1-h_0'+\fr{2}{r}h_0$, and substituting it back into \eqref{Lag3} results in \eqref{Lag2}.
From the new Lagrangian \eqref{Lag3}, one obtains the EOMs for $h_0$ and $h_1$, which can be solved in terms of $\chi$ as
	\be
	\begin{split}
	h_0&=-\fr{2r^2a_3a_4\dot{\chi}+4ra_2\brb{r(a_3\chi)'+2a_3\chi}}{4a_2\brb{r^2a_1-2(ra_3)'}-r^2a_4{}^2},\\
	h_1&=\fr{4a_3\dot{\chi}\brb{r^2a_1-2(ra_3)'}+2ra_4\brb{r(a_3\chi)'+2a_3\chi}}{4a_2\brb{r^2a_1-2(ra_3)'}-r^2a_4{}^2}.
	\end{split} \label{h0h1}
	\ee
Then the resubstitution into \eqref{Lag3} yields the following Lagrangian for $\chi$:
	\be
	\fr{2\ell+1}{2\pi}\mL^{(2)}=\fr{\ell(\ell+1)}{2(\ell-1)(\ell+2)}\sqrt{\fr{B}{A}}\brb{b_1\dot{\chi}^2-b_2\chi'{}^2+b_3\dot{\chi}\chi'-\ell(\ell+1)b_4\chi^2-V\chi^2}, \label{Lag4}
	\ee
where
	\be
	\begin{split}
	b_1&=\fr{r^2\mF\mH^2}{A\mF\mG+B\mJ^2}, \\
	b_2&=\fr{r^2AB\mG\mH^2}{A\mF\mG+B\mJ^2}, \\
	b_3&=\fr{2r^2B\mH^2\mJ}{A\mF\mG+B\mJ^2}, \\
	b_4&=\mH.
	\end{split} \label{bcoeff}
	\ee
Note that one cannot rewrite the Lagrangian~\eqref{Lag3} as \eqref{Lag4} for $\ell=1$ since the denominators of $h_0$ and $h_1$ in \eqref{h0h1} vanish in such a case.
The detailed expression of the potential~$V(r)$ is given in Appendix~\ref{appB}.

\subsection{Stability of $\ell\ge2$ modes}\label{opertmultista}
As was pointed out in \cite{Ogawa:2015pea}, for BH solutions (if they exist) to be stable, it is necessary that
	\be
	b_1>0,~~~b_2>0,~~~b_4>0. \label{bstability}
	\ee
The first condition guarantees the positive kinetic energy, while the latter two ensure the positive gradient energy.
From \eqref{bcoeff}, the criterion~\eqref{bstability} is equivalent to
	\be
	\mF>0,~~~\mG>0,~~~\mH>0. \label{stability}
	\ee
Now, let us investigate the behavior of these functions near the horizon.
Collecting terms that are potentially divergent in the vicinity of the horizon where $A\simeq0$, we obtain
	\be
	\mF\mG\approx-\fr{q^4}{A^2}\bra{2G_{4X}+\fr{A'}{A}\fr{X}{\psi'}G_{5X}}^2=-\bra{\fr{q\mJ}{A\psi'}}^2. \label{fghor}
	\ee
Provided that $\mJ\ne0$, this quantity is always negative, which implies the system is plagued by ghost or gradient instability (for the detailed arguments, see \cite{Ogawa:2015pea}).
Conversely, to avoid the instability, $\mJ$ must vanish at least in the vicinity of the horizon.
Such a situation is easily achieved if we choose $G_4$ and $G_5$ to be constant.
Without loss of generality, one can set $G_5=0$ since $\mL_5$ becomes total derivative for constant $G_5$.
Note that this class includes the solutions in \cite{Babichev:2016fbg} as a special case.

Let us now restrict our linear perturbation analysis to this particular class; i.e., we start from the following action:
	\be
	S'=\int d^4x\sqrt{-g}\left[ G_2(X)-G_3(X)\Box\phi+G_4R\right],
	\ee
where $G_4$ is a constant.
In this case, the functions in \eqref{fghj} become $\mF=\mG=\mH=2G_4$ and $\mJ=0$, and thus $G_4$ must be positive in order to satisfy \eqref{stability}.
Moreover, the potential given in Appendix~\ref{appB} drastically simplifies as
	\be
	V(r)=\fr{G_4}{A^2}\left[-r^2A'^2B+r^2A(A'B'+A''B)-A^2(4-4B+4rB'+r^2B'')\right]. \label{pot}
	\ee
Then, the EOM derived from \eqref{Lag4} becomes
	\be
	\fr{\pa}{\pa r}\bra{\fr{r^2B^{3/2}}{A^{1/2}}\chi'}-\fr{r^2B^{1/2}}{A^{3/2}}\ddot{\chi}-\sqrt{\fr{B}{A}}\left[ \ell(\ell+1)+\fr{V}{2G_4}\right]\chi=0. \label{chiEOM}
	\ee
In terms of the tortoise coordinate $r_\ast$ defined by $\fr{\pa}{\pa r}=\fr{1}{\sqrt{AB}}\fr{\pa}{\pa r_\ast}$ and a new variable~$\Psi\equiv \sqrt{\fr{B}{A}}\,r\chi$, \eqref{chiEOM} simplifies as
	\begin{align}
	&\fr{\pa^2\Psi}{\pa r_\ast^2}-\ddot{\Psi}-V_{\rm eff}\Psi=0, \label{omeq1} \\
	&V_{\rm eff}(r)\equiv A\fr{(\ell-1)(\ell+2)}{r^2}+\fr{2AB}{r^2}-\fr{(AB)'}{2r}.
	\end{align}
Note that \eqref{omeq1} should be regarded as a differential equation with respect to $(t,r_\ast)$.
For the Schwarzschild solution with $A=B=1-\fr{r_h}{r}$, the effective potential $V_{\rm eff}(r)$ coincides with the well-known Regge-Wheeler potential~\cite{Regge:1957td}.
If we focus on a mode with frequency $\omega$, \eqref{omeq1} is written in the form of an eigenvalue equation:
	\be
	\hat{H}\Psi=\omega^2\Psi,~~~\hat{H}\equiv-\fr{d^2}{dr_\ast^2}+V_{\rm eff}.
	\ee
If all the eigenvalues of $\hat{H}$ are positive, the solution is stable for the fixed mode.\footnote{Note that we only consider perturbations whose support is compact on the initial surface.}
However, this does not necessarily guarantee the linear stability of all the solutions evolved from the regular initial data.
To prove the linear stability, one has to address the boundedness and asymptotic behavior of the general solutions~\cite{Dafermos:2016uzj}.
Since this problem is beyond the scope of this paper, here we only focus on the mode analysis.

The positivity of the eigenvalues is equivalent to
	\be
	\langle\varphi,\hat{H}\varphi\rangle\equiv \int dr_\ast\brb{\bra{\fr{d\varphi}{dr_\ast}}^2+V_{\rm eff}\varphi^2}>0 \label{potsta}
	\ee
for any function~$\varphi$ with compact support.
Here, the integration range runs over all the values of $r_\ast$ corresponding to the possible range of $r$.
To prove \eqref{potsta}, the so-called $S$-deformation technique is useful \cite{Ishibashi:2003ap,Kodama:2003kk}.
Let us introduce the deformed differential operator and potential as
	\begin{align}
	\ti{D}&\equiv \fr{d}{dr_\ast}+S, \\
	\ti{V}&\equiv V_{\rm eff}+\fr{dS}{dr_\ast}-S^2,
	\end{align}
with $S$ being an arbitrary function.
Note that \eqref{potsta} can be rewritten in terms of $\ti{D}$ and $\ti{V}$ as
	\be
	\langle\varphi,\hat{H}\varphi\rangle=\int dr_\ast\brb{\bigl(\ti{D}\varphi\bigr)^2+\ti{V}\varphi^2}>0.
	\ee
Therefore, if one manages to find $S$ such that $\ti{V}>0$ for all $r>r_h$, then it completes the proof of the mode stability.
In our present case, we choose $S=\fr{\sqrt{AB}}{r}$ to obtain
	\be
	\ti{V}=A\fr{(\ell-1)(\ell+2)}{r^2}>0,
	\ee
since we are now working on $\ell\ge 2$.
Thus, the odd-parity perturbations of BH solutions in the case of constant $G_4$ and $G_5$ are fully stable for fixed modes.
Note that we did not assume any specific form of the background solution.

Besides the above case, there are still other possibilities that can avoid the ghost and gradient instabilities.
Since we could not figure out them all, here we just provide a single example of such possibilities.
Let us consider the following subclass of shift-symmetric Horndeski theories:
	\be
	\begin{split}
	G_2(X)=g_2+f_2(X),~~~&f_2(0)=f_{2X}(0)=0,\\
	G_3(X)=g_3+f_3(X),~~~&f_3(0)=f_{3X}(0)=0,\\
	G_4(X)=g_4+f_4(X),~~~&f_4(0)=f_{4X}(0)=f_{4XX}(0)=0,\\
	G_5(X)=g_5+f_5(X),~~~&f_5(0)=f_{5X}(0)=0,\\
	\end{split} \label{ex}
	\ee
where $g_i$ are constant.\footnote{Since the constant parts of $G_3$ and $G_5$ just result in total derivative in the Lagrangian~\eqref{shiftsymact}, one can set $g_3=g_5=0$ from the beginning.}
In this case, one can easily check that the following configuration satisfies the background EOMs given in Appendix~\ref{appA}:
	\be
	A(r)=B(r)=1-\fr{2M}{r}+\fr{g_2}{6g_4}r^2,~~~X(r)=0, \label{bgsol}
	\ee
where $M$ is a constant.\footnote{Although \eqref{bgsol} satisfies the background EOMs, there is still another consideration that theories of the type~\eqref{ex} may not be well posed as an initial value problem \cite{Graham:2014ina}.}
The configuration of the scalar field is obtained from $X=0$ as
	\be
	\phi(t,r)=q\bra{t+\int^r\!\!\!\fr{d\ti{r}}{A(\ti{r})}}. \label{scasol}
	\ee
Since $\mF=\mG=\mH=2g_4$ for this solution, the criterion~\eqref{stability} for the absence of ghost/gradient instability is satisfied if $g_4>0$.
Furthermore, since the potential takes exactly the same form as in \eqref{pot}, no other instability arises for the solution~\eqref{bgsol} in the same way as the above case of $G_4$ and $G_5$ being constant.
In other words, the degeneracy with GR still remains even at the odd linear perturbation level.
To tell the difference between them, one has to examine even-parity perturbations or proceed to nonlinear level.

\subsection{Dipole modes}\label{opertdi}
As was shown in \cite{Kobayashi:2012kh,Ogawa:2015pea}, the dipole perturbations are related with the slow rotation of a BH.
Since the structure of the Lagrangian is the same as in \cite{Ogawa:2015pea}, we can follow the same arguments to clarify the physical meaning of the dipole perturbations.

We start from the Lagrangian~\eqref{Lag2} with $\ell=1$.
Since the coefficients given in \eqref{acoeff} satisfy
	\be
	a_1=\fr{2(ra_3)'}{r^2},~~~a_2=a_4=0
	\ee
for $\ell=1$, \eqref{Lag2} simplifies as
	\be
	\fr{3}{2\pi}\mL^{(2)}=a_3\bra{\dot{h}_1-h_0'+\fr{2}{r}h_0}^2. \label{Lagdipole}
	\ee
We eliminate $h_1$ by choosing the gauge function $\Lambda$ appropriately.
Note that, as can be read off from \eqref{gaugetrnsf}, there still remains gauge DOF such that $\Lambda=c(t)r^2$.
Then the EOMs derived from \eqref{Lagdipole} become
	\be
	\begin{split}
	h''_0+\fr{a_3'}{a_3}h'_0-\fr{2(ra_3)'}{r^2a_3}h_0&=0,\\
	\dot{h}'_0-\fr{2}{r}\dot{h}_0&=0.
	\end{split} \label{dipoleeom2}
	\ee
As in \cite{Ogawa:2015pea}, the general solution to this system of equations is written as
	\be
	h_0=\bar{c}(t)r^2+\fr{3Jr^2}{4\pi}\int^r\fr{d\ti{r}}{\ti{r}^4a_3(\ti{r})}.
	\ee
Here, the first term is just a gauge mode and thus can be eliminated by use of the residual gauge DOF with $c(t)=\int^td\ti{t}\,\bar{c}(\ti{t})$.
On the other hand, the second term represents the slow rotation of a BH, and the integration constant $J$ corresponds to the BH angular momentum.
Indeed, if $a_3$ is constant (which is the case for the two classes of solutions in \S \ref{opertmultista} that are mode stable under odd-parity perturbations), then we obtain
	\be
	h_0=-\fr{J}{4\pi a_3r}. \label{fdf}
	\ee
In the case of GR where $a_3=\Mpl^2$, \eqref{fdf} coincides with the $t\varphi$-component, namely, the frame-dragging function of the Kerr metric up to first order in the angular momentum.

Next, let us consider the following choice of the four arbitrary functions in \eqref{shiftsymact}:
	\be
	G_2(X)=X,~~~G_3(X)=0,~~~G_4(X)=\fr{\Mpl^2}{2},~~~G_5(X)=-4\alpha \ln X.
	\ee
It is known that this choice is equivalent to the following Einstein-dilaton-Gauss-Bonnet action \cite{Kobayashi:2011nu}:
	\be
	S_{\rm EdGB}=\int d^4x\sqrt{-g}\bra{\fr{\Mpl^2}{2}R-\fr{1}{2}(\pa_\mu\phi)^2+\alpha \phi R_{\rm GB}^2},
	\ee
which includes a linear coupling of the scalar field to the Gauss-Bonnet invariant~$R_{\rm GB}^2\equiv R^2-4R_{\mu\nu}R^{\mu\nu}+R_{\mu\nu\la\si}R^{\mu\nu\la\si}$.
In this case, the coefficient $a_3$ is no longer constant, and thus, the frame-dragging function is different from that in GR.
These results are consistent with those of \cite{Maselli:2015yva}, in which solutions were obtained at first order in the Hartle-Thorne slow-rotation approximation~\cite{Hartle:1967he,Hartle:1968si}.

\section{Conclusions}\label{conclusion}

In the present paper, we obtained the quadratic action for odd-parity perturbations of BHs with linearly time-dependent scalar hair in shift-symmetric Horndeski theories and performed the mode analysis.
It turned out that the BHs are plagued by ghost or gradient instability for a wide class of theories, extending the result for the reflection-symmetric subclass in \cite{Ogawa:2015pea}.
We also derived the conditions to evade such instabilities and presented two particular classes of BH solutions belonging to this case:~The first consists of solutions for constant $G_4$ and $G_5$, while the second is the solution~\eqref{bgsol} under the assumption~\eqref{ex}.
Furthermore, these classes are fully stable in the sense of mode stability under linear odd-parity perturbations.
We expect there are still other physically possible solutions.

Note that, what we analyzed in the present paper is the mode stability.
In general, one cannot conclude a solution is linearly stable only by its mode stability~\cite{Dafermos:2016uzj}.
However, the mode analysis is still a useful method to prove the existence of an instability.
In a subsequent paper, we investigate the mode stability of even-parity perturbations for solutions that evade the instability at the level of odd modes.


\acknowledgements{
This work was supported in part by
JSPS Grant-in-Aid for Young Scientists (B) No.~15K17632 (T.\,S.), 
MEXT Grant-in-Aid for Scientific Research on Innovative Areas ``New Developments in Astrophysics Through Multi-Messenger Observations of Gravitational Wave Sources'' No.~15H00777 (T.\,S.) and ``Cosmic Acceleration'' No.~15H05888 (T.\,S.).
}


\appendix
\section{Background equations}\label{appA}

Among the five background EOMs~$\mE_A,\mE_B,\mE_C,\mE_D$ and $\mE_\psi$, the independent components are only $\mE_A,\mE_B,\mE_C$.
We decompose them into terms with and without the scalar velocity charge $q$ as in \cite{Maselli:2015yva,Ogawa:2015pea}:
	\be
	\mE_\Phi\equiv\mE_\Phi^{(0)}+\fr{q^2}{A}\mE_\Phi^{(t)},~~~(\Phi=A,B,C),
	\ee
where
	\begin{align}
	\mE_A^{(0)}&=G_2-\fr{1}{2}B\psi'{}^2(B'\psi'+2B\psi'')G_{3X}-\fr{2}{r}\bra{B'+\fr{B-1}{r}}G_4-\fr{2B^2\psi'}{r}\bra{\fr{\psi'}{r}+\fr{2B'}{B}\psi'+2\psi''}G_{4X} \nonumber \\
	&~~~+\fr{2B^3\psi'{}^3}{r}\bra{\fr{B'}{B}\psi'+2\psi''}G_{4XX}+\fr{B^2\psi'{}^2}{2r^2}\bra{(5B-1)\fr{B'}{B}\psi'+2(3B-1)\psi''}G_{5X}-\fr{B^4\psi'{}^4}{2r^2}\bra{\fr{B'}{B}\psi'+2\psi''}G_{5XX}, \\
	\mE_A^{(t)}&=-G_{2X}+B\bra{\fr{2\psi'}{r}+\fr{B'}{2B}\psi'+\psi''}G_{3X}+\fr{2}{r}\bra{B'+\fr{B-1}{r}}G_{4X}-\fr{2B^2\psi'}{r}\bra{\fr{\psi'}{r}+\fr{B'}{B}\psi'+2\psi''}G_{4XX} \nonumber \\
	&~~~-\fr{B}{2r^2}\bra{(3B-1)\fr{B'}{B}\psi'+2(B-1)\psi''}G_{5X}+\fr{B^3\psi'{}^2}{2r^2}\bra{\fr{B'}{B}\psi'+2\psi''}G_{5XX}, \\
	\mE_B^{(0)}&=G_2+B\psi'{}^2G_{2X}-\fr{B^2\psi'{}^3}{2}\bra{\fr{4}{r}+\fr{A'}{A}}G_{3X}-\fr{2}{r}\bra{\fr{A'}{A}B+\fr{B-1}{r}}G_4-\fr{2B\psi'{}^2}{r}\bra{\fr{2A'}{A}B+\fr{2B-1}{r}}G_{4X} \nonumber \\
	&~~~+\fr{2B^3\psi'{}^4}{r}\bra{\fr{1}{r}+\fr{A'}{A}}G_{4XX}+\fr{B^2\psi'{}^3}{2r^2}\fr{A'}{A}(5B-1)G_{5X}-\fr{B^4\psi'{}^5}{2r^2}\fr{A'}{A}G_{5XX}, \\
	\mE_B^{(t)}&=\fr{A'}{A}\bra{\fr{B\psi'}{2}G_{3X}+\fr{2B}{r}G_{4X}-\fr{2B^2\psi'{}^2}{r}G_{4XX}-\fr{B\psi'}{2r^2}(3B-1)G_{5X}+\fr{B^3\psi'{}^3}{2r^2}G_{5XX}}, \\
	\mE_C^{(0)}&=G_{2X}-\fr{B\psi'}{2}\bra{\fr{4}{r}+\fr{A'}{A}}G_{3X}-\fr{2}{r}\bra{\fr{A'}{A}B+\fr{B-1}{r}}G_{4X}+\fr{2B^2\psi'{}^2}{r}\bra{\fr{1}{r}+\fr{A'}{A}}G_{4XX}+\fr{B\psi'}{2r^2}\fr{A'}{A}(3B-1)G_{5X} \nonumber \\
	&~~~-\fr{B^3\psi'{}^3}{2r^2}\fr{A'}{A}G_{5XX}, \\
	\mE_C^{(t)}&=\fr{A'}{A\psi'}\bra{\fr{1}{2}G_{3X}-\fr{2B\psi'}{r}G_{4XX}-\fr{B-1}{2r^2}G_{5X}+\fr{B^2\psi'{}^2}{2r^2}G_{5XX}}.
	\end{align}
Here, the overall factors are chosen so that the coefficients of $G_2$ or $G_{2X}$ in $\mE_a^{(0)}$ become unity.
Note also that we have set $C=0$ and $D=1$ after deriving the EOMs.

The remaining EOMs~$\mE_D$ and $\mE_\psi$ can be written in terms of $\mE_A,\mE_B,\mE_C$ as
	\begin{align}
	\mE_D&\equiv\fr{1}{r^2}\sqrt{\fr{B}{A}}\fr{\delta S}{\delta D}=-\fr{rA'}{4A}\mE_A+\bra{1+\fr{rA'}{4A}}\mE_B+\fr{r}{2}\mE_B'-\sqrt{\fr{B}{A}}\fr{\psi'}{2r}\fr{d}{dr}\bra{\sqrt{AB}r^2\psi'\mE_C}, \\
	\mE_\psi&\equiv\sqrt{\fr{B}{A}}\fr{\delta S}{\delta \phi}=\sqrt{\fr{B}{A}}\fr{d}{dr}\bra{\sqrt{AB}r^2\psi'\mE_C}.
	\end{align}


\section{Expression of $V(r)$}\label{appB}
The potential $V(r)$ defined in \eqref{Lag4} is written in terms of the functions in \eqref{fghj} as
	\begin{align}
	V(r)&=\fr{\mH}{2(A\mF\mG+B\mJ^2)^2}\boldsymbol{\Bigl\{}\!-B\boldsymbol{\Bigl[}r^2A'\mG\mH(A'\mF\mG+B'\mJ^2) \nonumber \\
	&~~~+B\mJ\{2r^2A'\mG\mH\mJ'+r\mJ[\mG(-rA''\mH+rA'\mH'+4A'\mH)-rA'\mG'\mH]+4\mJ^3\}\boldsymbol{\Bigr]} \nonumber \\
	&~~~-A^2\mG^2\boldsymbol{\Bigl[}-r\mF'[rB'\mH+2B(r\mH'+2\mH)]+4\mF^2 \nonumber \\
	&~~~~~~~~~~~~~~~+\mF[r(rB''\mH+3rB'\mH'+4B'\mH)+2B(r^2\mH''+2r\mH'-2\mH)]\boldsymbol{\Bigr]} \nonumber \\
	&~~~+A\boldsymbol{\Bigl[}r^2B'\mG\mH(A'\mF\mG+B'\mJ^2)+B\{\mF\mG[r^2\mG(A''\mH+A'\mH')-8\mJ^2] \nonumber \\
	&~~~~~~~~~~-r^2[A'\mF'\mG^2\mH+B'\mG'\mH\mJ^2+\mG\mJ(B''\mH\mJ+B'\mH'\mJ-2B'\mH\mJ')]\} \nonumber \\
	&~~~~~~~~~~+2B^2\mJ\boldsymbol{\Bigl(}\mG\{r[2r\mH'\mJ'-\mJ(r\mH''+2\mH')]+2\mH(2r\mJ'+\mJ)\}-r\mG'\mJ(r\mH'+2\mH)\boldsymbol{\Bigr)}\boldsymbol{\Bigr]}\boldsymbol{\Bigr\}}.
	\end{align}


\bibliography{BH-odd}

\end{document}